\documentclass{iau-JDSS}
\usepackage{graphicx}

\title[Obscured and distant clusters] 
{SpS5 - III. Matter ejection and feedback}

\author[short author list]   
{Ya\"el Naz\'e$^1$
  \thanks{FNRS Research Associate},
  \and Xiao Che$^2$, Nick L.J. Cox$^3$, Jos\'e H. Groh$^4$, Martin Guerrero$^5$, Pierre Kervella$^6$, Chien-De Lee$^7$, Mikako Matsuura$^8$, Sally Oey$^2$, Guy S. Stringfellow$^9$, Stephanie Wachter$^{10}$}

\affiliation{$^1$Department AGO, All\'ee du 6 Ao\^ut 17, B5C, 4000-Li\`ege, Belgium, email: naze@astro.ulg.ac.be\\[\affilskip]
$^2$ Dpt of Astronomy, Univ. of Michigan, 830 Dennison Bldg, Ann Arbor, MI 48109-1042, USA\\
$^3$Institute of Astronomy, KU Leuven, Celestijnenlaan 200D, Leuven, Belgium\\
$^4$ Geneva Observ., Univ. of Geneva, Chemin des Maillettes 51, CH-1290 Sauverny, Switzerland\\
$^5$ IAA-CSIC, Glorieta de la Astronomia s/n, 18008 Granada, Spain \\
$^6$ LESIA, Obs. de Paris, CNRS UMR 8109, UPMC, 5 place J. Janssen, 92195, Meudon, France\\
$^7$ Graduate institute of astronomy, National central university, Taiwan\\
$^8$ Department of Physics and Astronomy, UCL, Gower Street, London WC1E 6BT, UK \\
$^9$ Center for Astroph. and Space Astr., Univ. of Colorado, 389 UCB,  Boulder, CO, USA\\
$^{10}$ IPAC, California Inst. of Technology, Pasadena, CA 91125, USA
}

\pubyear{2012}
\volume{Volume 16}  
\pagerange{20--29}
\date{?? and in revised form ??}
\jname{Highlights of Astronomy, Volume 16}
\editors{T. Montmerle, ed.}
\begin{document}

\maketitle

\begin{abstract}
The last part of SpS5 dealt with the circumstellar environment. Structures are indeed found around several types of massive stars, such as blue and red supergiants, as well as WRs and LBVs. As shown in the last years, the potential of IR for their study is twofold: first, IR can help discover many previously unknown nebulae, leading to the identification of new massive stars as their progenitors; second, IR can help characterize the nebular features. Current and new IR facilities thus pave the way to a better understanding of the feedback from massive stars.

\end{abstract}

\firstsection 
\section{Introduction}

Circumstellar material holds clues about the mass-loss history of massive stars. Indeed, as the winds interact with the interstellar medium (wind-blown bubbles, bow shocks), they leave a characteristic signature that depends on the wind properties. Moreover, the material ejected during short eruptive phases is visible as nebulae around massive stars. The analysis of these features reveals which material was ejected and in which quantity. With the recent reduction in mass-loss rates, these episodes of enhanced mass-loss have gained more attention, as they seem more crucial than ever in the evolution of massive stars.

Another reason to study the close environment of massive stars is to better understand the evolution of supernova remnants (SNRs). Indeed, the famous rings of SN1987A may only be understood if one considers the previous mass-loss episodes of the progenitor. Morphology is not the only SNR parameter which is affected, as the SNR dynamics in an homogeneous medium or in winds and circumstellat ejecta is not identical.

For its study, the IR provides several key diagnostics. Continuum emission in this range is provided by heated dust, which may have a range of temperatures depending of the framework (very close hot features, large, old, and cool bubbles). In addition, IR lines probe the many phases of the material: molecules (e.g. PAHs) for the neutral material, ionized metals for HII regions,... 

This summary of SpS5 - part III examines each case of circumstellar environment in turn, and concludes with the potential offered by current and future facilities.

\section{Blue supergiants}
Circumstellar structures around BSGs have been predominantely identified as bow shocks around runaway stars. Originally discovered with IRAS (e.g. Van Buren \& McCray, 1988, ApJ, 329, L93), such structures have also been seen with MSX and WISE (Peri et al. 2012). A more general survey of BSGs, i.e. not targeting runaway stars, with objects selected from Crowther et al. (2006) and Przybilla et al. (2010), reveals IR material around six of the 45 targets at 22$\mu$m with WISE, also mostly in the form of bow shocks (Wachter, in prep). Several examples of bipolar nebulae around BSGs are also known (e.g. Sher 25, Smartt et al. 2002; HD 168625, Smith 2007). However, this material could have also been ejected during an LBV phase, since LBVs can exhibit BSG spectra, and we will therefore concentrate on the bow shocks.

\begin{figure}
\centering
\includegraphics[width=6cm]{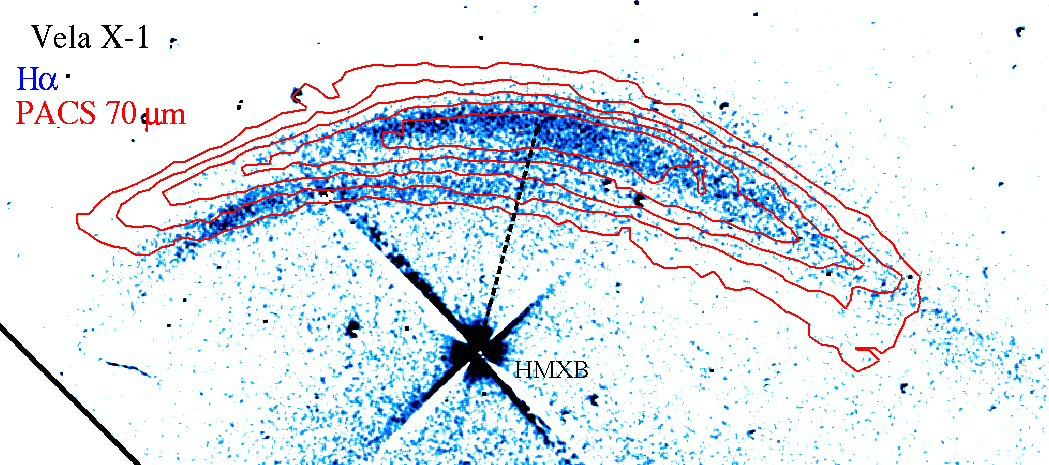}
\includegraphics[width=6cm]{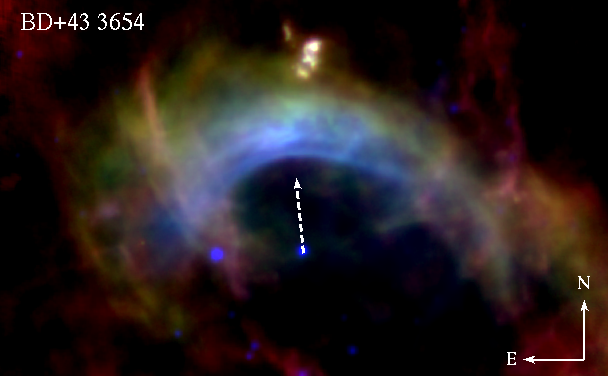}
\caption{{\it Left:} H$\alpha$ emission (greyscale) of Vela X-1 with PACS 70$\mu$m emission contours shown on top. {\it Right:} Colour composite image of bow shock of BD+43$^{\circ}$3654 (WISE 12$\mu$m in blue, PACS 70$\mu$m in green, and PACS 160$\mu$m in red). The direction of proper motion is indicated by the arrow in both cases. From Cox et al. (in prep.).}
\label{Cox}
\end{figure}

Runaway stars have large stellar velocities (above 30\,km\,s$^{-1}$) resulting from dynamical interactions in (dense) clusters or from a supernova explosion in a binary system. These stars can thus travel at supersonic speeds through the local medium giving rise to ``bow shocks'' as their stellar winds interact with the surrounding medium, which has been previously ionised by stellar photons from the hot star (Weaver 1977). The occurrence of such bow shocks has been shown to depend primarily on the ISM conditions (Huthoff \& Kaper 2002). For example, even a runaway star may travel at subsonic speeds in the tenuous interior of a superbubble, where the sound speed can be as much as 100\,km\,s$^{-1}$, hence no (detectable) bow shock will be produced in that case. The filling factor of ISM with $v_\mathrm{sound} \leq 10$\,km\,s$^{-1}$ is 20\% and 75\% of O-stars have velocities $\geq$10\,km\,s$^{-1}$, so the expected fraction of O-stars with bow shocks is $\sim$15\%. This is remarkably similar to the values derived from IRAS and WISE observations (Noriega-Crespo et al. 1997, Peri et al. 2012).

Once formed, the size, shape and morphology of a bow shock depends on both stellar (wind kinetic energy and stellar velocity) and interstellar parameters (density and temperature). In particular the ratio $v_\star/v_\mathrm{wind}$ indicates whether or not instabilities are likely to develop (Dgani et al. 1996), and the stand-off distance between the star and the apex of the shock is determined from the pressure balance between the stellar wind and the ISM (see analytical results by Wilkin 1996 and simulations by e.g. Comeron \& Kaper 1998, Blondin \& Koerwer 1998). Independent estimates of the wind parameters can thus be inferred from bow shocks, which serves as a useful check for atmosphere models, but the values are sensitive to the ISM properties, which are not always known with precision. 

\begin{figure}
\includegraphics[width=14cm]{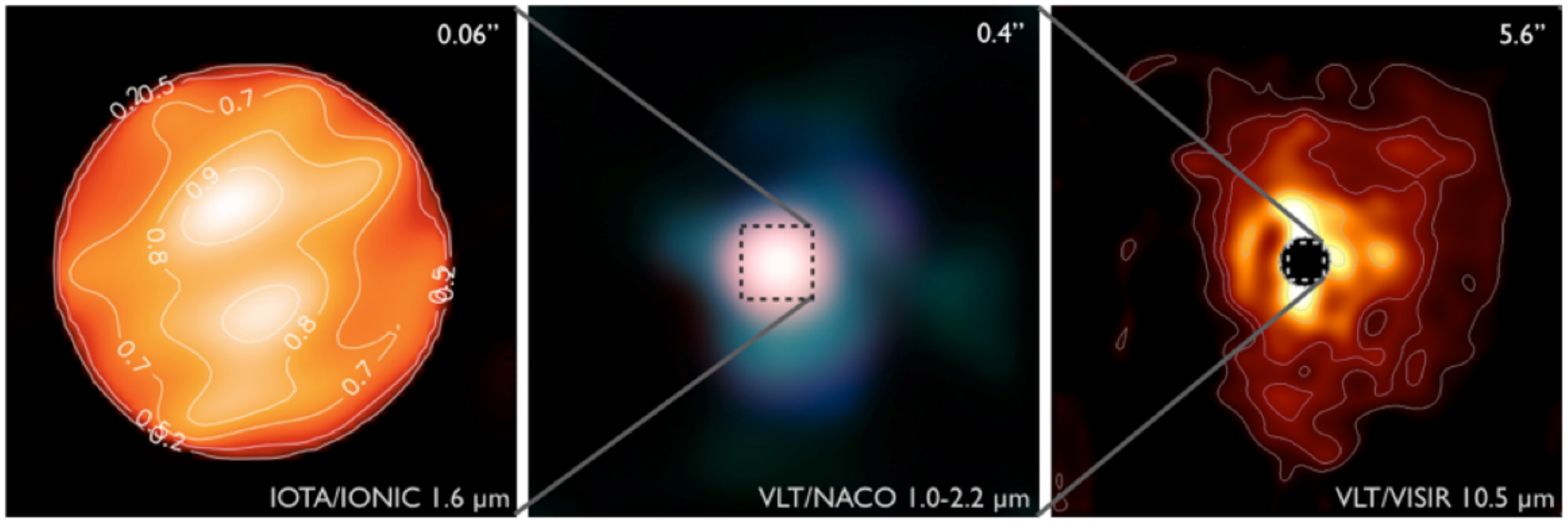}
\caption{{\it Left:} Interferometric image of the photosphere of Betelgeuse obtained by Haubois et al. (2009), showing its inhomogeneous surface brightness. {\it Center:} VLT/NACO adaptive optics tricolor composite image (RGB=KHJ) obtained by Kervella et al. (2009), showing the emission from a compact molecular envelope. {\it Right:} VLT/VISIR image at $10.49\,\mu$m of the dust thermal emission obtained by Kervella et al. (2011). North is up, East to the left, and the field of view is given in the upper right corner of each image. }
\label{Betel}
\end{figure}

Currently, a small survey with Herschel-PACS of 5 runaways with known bow-shocks is ongoing: $\alpha$\,Cam, $\zeta$\,Oph, $\tau$\,CMa, Vela X-1 and BD+43$^{\circ}$3654 (Cox et al., in preparation). For Vela X-1, the peak emission of the dust emission is co-spatial with the most prominent H$\alpha$ arc seen in the supposed direction of space motion (Fig. \ref{Cox}): it is concluded that the outer shock is radiative, but the inner shock is adiabatic, though some H$\alpha$ emission possibly related to (part of) the inner termination shock is also detected. From the analysis of its ``puffed-up'' bow shock (Fig. \ref{Cox}), the mass-loss rate of BD+43$^{\circ}$3654 (O4If) was found to be 10$^{-4}$\,M$_{\odot}$\,yr$^{-1}$: this is very high (by 2 orders of magnitude) in view of current mass-loss rate estimates of such stars, but the exact value strongly depends on ISM density, which need to be refined. The dust temperature, $\sim$~45~K, is compatible with heating by stellar photons only, suggesting there is no additional shock-heating of grains. The thickness of a bow shock ($\sim$~1~pc) suggests a Mach number close to unity, implying a ISM temperature of 10$^3$ -- 10$^4$~K. 

\section{Red supergiants}
Circumstellar structures on scales of a few arcseconds or less around RSGs have been revealed through interferometric techniques (e.g. Monnier et al. 2004). Stencel et al. (1988, 1989) reported the IRAS detection of resolved shells with typical radii of a few arcminutes around RSGs for a significant fraction (25\%) of their sample. However, higher resolution Spitzer images fail to confirm several of these extended structures (Wachter, in prep), indicating that a systematic survey is needed to ascertain the occurrence of large scale circumstellar shells around RSGs. 

\begin{figure}
\begin{center}
\includegraphics[width=12cm]{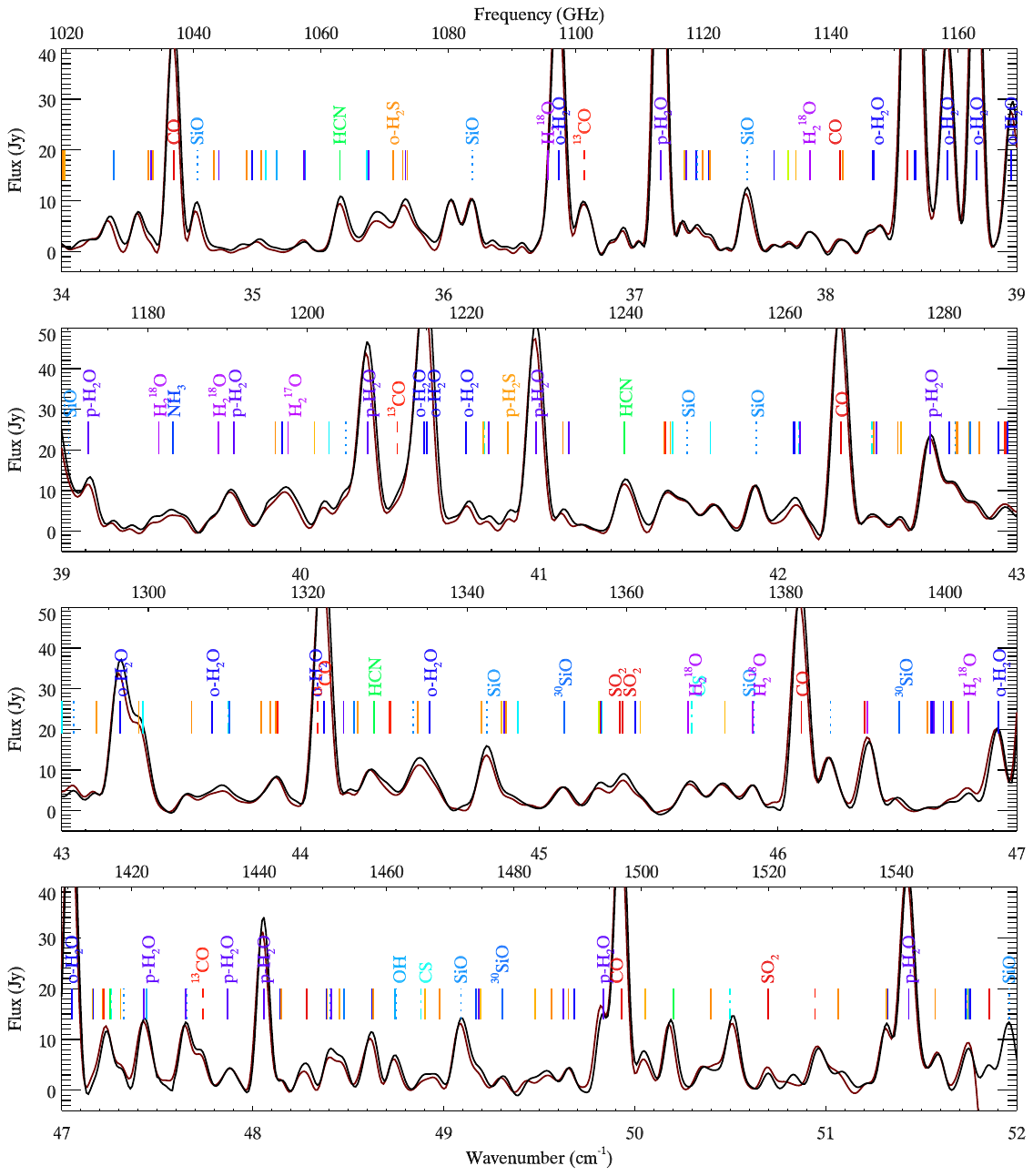}
\caption{Continuum subtracted Herschel SPIRE spectrum of VY CMa from 294$\mu$m to 192$\mu$m. Multitude of molecular lines have been detected. From Matsuura et al. (in prep.).}
\label{Mik}
\end{center}
\end{figure}

A few (famous) cases have however been studied in depth. One of these is Betelgeuse, a cool (3600K), large (700 R$_{\odot}$), rather massive (10--15 M$_{\odot}$), luminous ($>10^5$ L$_{\odot}$), and nearby (150 pc) star. Because of its distance, Betelgeuse can be probed on almost all scales, providing a unique panorama of stellar surroundings (Fig. \ref{Betel}). Space-based and interferometric instruments (e.g. HST, IOTA/Ionic and VLTI/Pionier) revealed the photosphere, notably the expected non-uniformities due to large convection cells. Adaptive optics imaging in the near-IR (e.g. NACO) and (radio or IR) interferometers unveiled the properties of the internal, compact molecular envelope (1--10 R$_*$). Precursors of dust have been found there, as well as an extended ``plume'' reaching 6R$_*$ and maybe linked to a hot spot on the photosphere. High-res imaging (e.g. VLT/VISIR) shows the envelope at intermediate scales (10--100 R$_*$), where the dust forms (a possible signature of silicates has been found). At these small and intermediate scales, Betegeuse presents a complex circumstellar envelope (with knots and filaments) at all wavelengths, which implies an inhomogeneous spatial distribution of the material lost by the star. Finally, at the largest scale, IR imagers such as Herschel unveil the cool external envelope (100-10000 R$_*$), where a bow shock with the ISM is detected (Cox et al. 2012).

\begin{figure}
\begin{center}
\includegraphics[width=10cm]{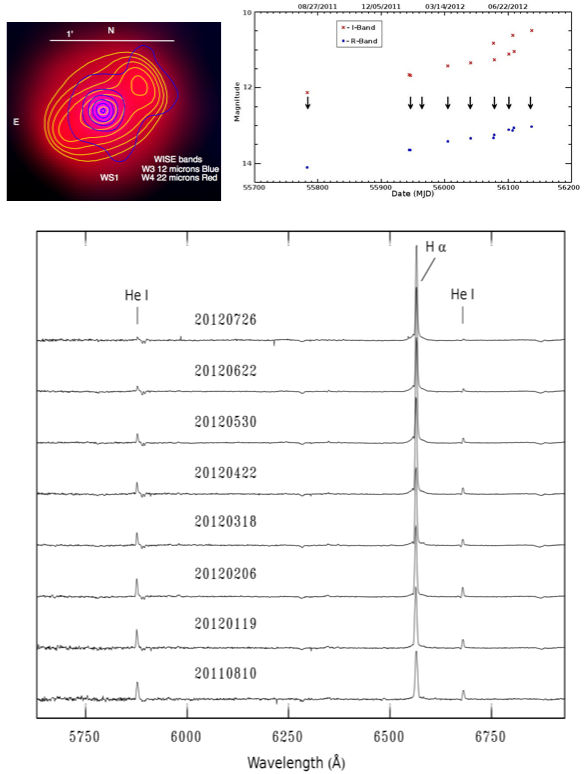}
\caption{{\it Top left:} A WISE color composite of 12 $\mu$m (blue; blue contours) and 22 $\mu$m (red; yellow contours) of WS1, discovered and initially  characterized by Gvaramadze et al. (2012). The contours for each band help illuminate the morphology of the nebular material, which has an overall SE-NW elongation, reminiscent of bipolar structure. {\it Bottom left:} Optical photometric monitoring since discovery show both the R and I light curves have brightened by about 1 magnitude over the last year. Arrows indicate times when same night spectroscopy were secured. {\it Right panel:} Optical spectroscopic monitoring indicates evolution to cooler temperature with near disappearance of the He I 5876 $\rm \AA$ and 6678 $\rm \AA$ and changes in the $\rm H_\alpha$ line profile. Figures from Stringfellow et al. (in preparation).}
\label{Guy}
\end{center}
\end{figure}

Herschel has also probed the envelope of other red supergiants (Groenewegen et al. 2011). Turning in particular to the case of VY CMa (Matsuura et al., in prep.), the potential of IR spectroscopy is obvious. Herschel-SPIRE reveals a rich spectrum, with a dust continuum and hundreds of lines dues to molecules (one third linked to water, others to CO, CS, SiO,...), which constrain the envelope's properties. For example, the isotopic ratio $^{12}$C/$^{13}$C is found to be 6.5, in agreement with observations of other RSGs but at odds with theoretical predictions which are four times higher at least. Very strong emission of submm molecular lines can be explained if a temperature gradient is present in the envelope, e.g. because of dust formation at a certain radius. 

\section{Luminous Blue Variables}

Because of their spectacular eruptions, LBVs are the most well-known cases of massive stars with ejecta. It is not yet certain, however, at what stage (BSG? after a RSG phase or not?) this material is ejected, and how (multiple events?). LBVs are rare: in the list of Clark et al. (2005), there are only 12 confirmed and 23 candidate LBVs. IR has played a key role in recent years. The search, through surveys like MIPSGAL, of round-shaped nebulae with luminous central stars resulted in the discovery of many new nebulae: 62 shells in Wachter et al. (2010), 115 shells and bipolar nebulae in Gvaramadze et al. (2010), 416 structures in Mizuno et al. (2010). Many of these nebulae are preferentially detected with Spitzer 24$\mu$m band, indicating relatively cold material. 

Identifying shell-like structures is only the first step. To ascertain a cLBV status, the central object needs to be studied spectroscopically. This was done for many of these new detections (c.f., Gvaramadze et al. 2010; Wachter et al. 2010; Stringfellow et al. 2012a,b, and in preparation). The classification does not rely on the presence of a particular line, but rather on the morphological resemblance of the spectra to spectra of known LBVs - while not 100\% perfect (some peculiar O and WR stars display similar features), this method has the advantage of being simple and rather robust. A more definitive answer can be provided through photometric and/or spectroscopic monitorings. Indeed, as their name indicate, LBVs should be {\it variable}. Near-simultaneous photometric and spectroscopic monitoring in the optical (and IR) of about a dozen newly identified candidate LBVs has revealed that WS1 (discovered by Gvaramadze et al. 2012) is indeed a bona fide LBV, presently displaying what appears to be S Dor type variability as shown in Fig. \ref{Guy} (Stringfellow et al. 2012, in preparation).

\begin{figure}
\begin{center}
\includegraphics[width=12cm]{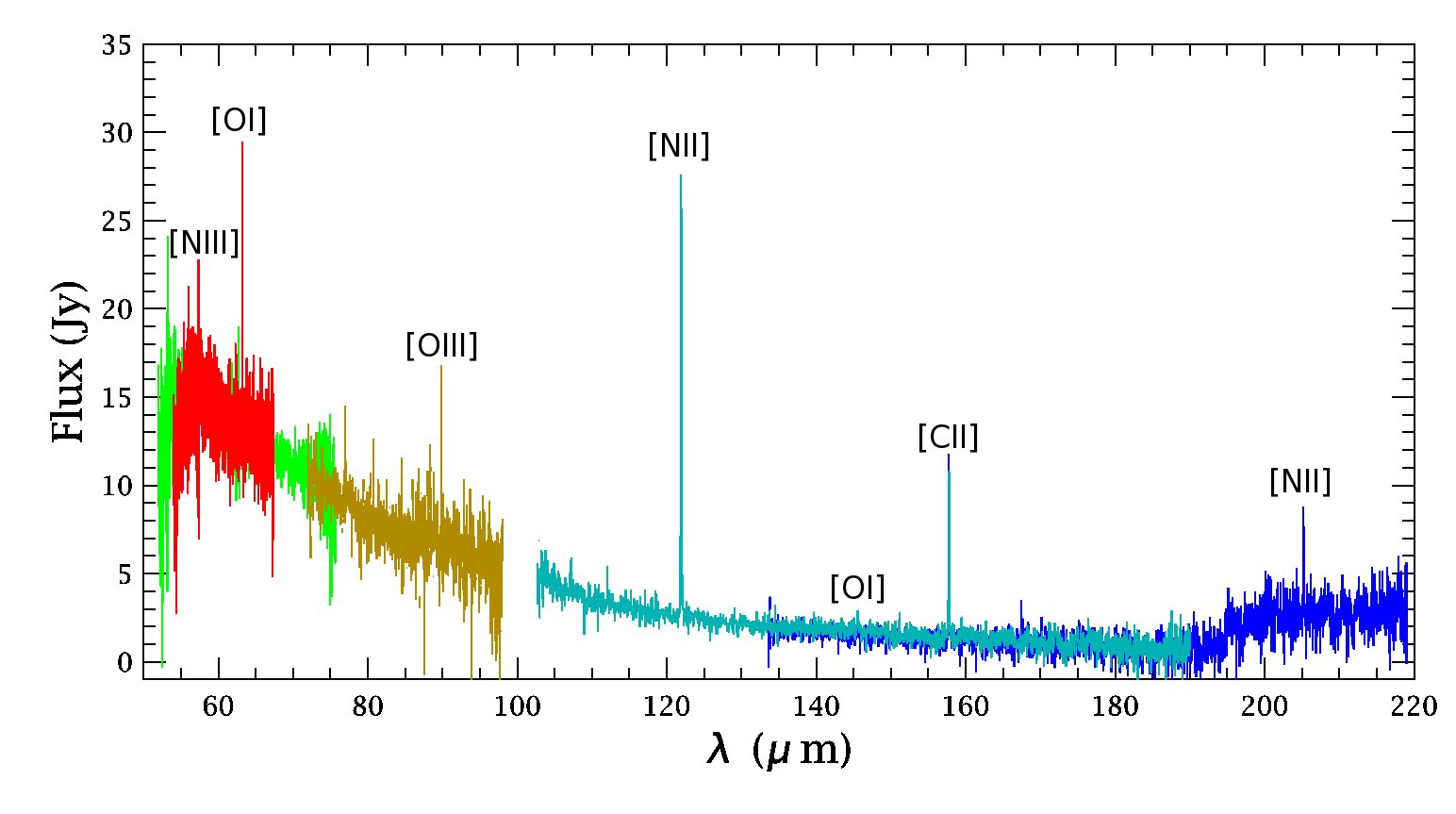}
\caption{PACS spectrum (central spaxel) of WRAY 15-751 nebula, showing the lines from the ionized and neutral regions (from Vamvatira-Nakou et al., submitted).}
\label{Chloi}
\end{center}
\end{figure}

\begin{figure}
\begin{center}
\includegraphics[width=4cm]{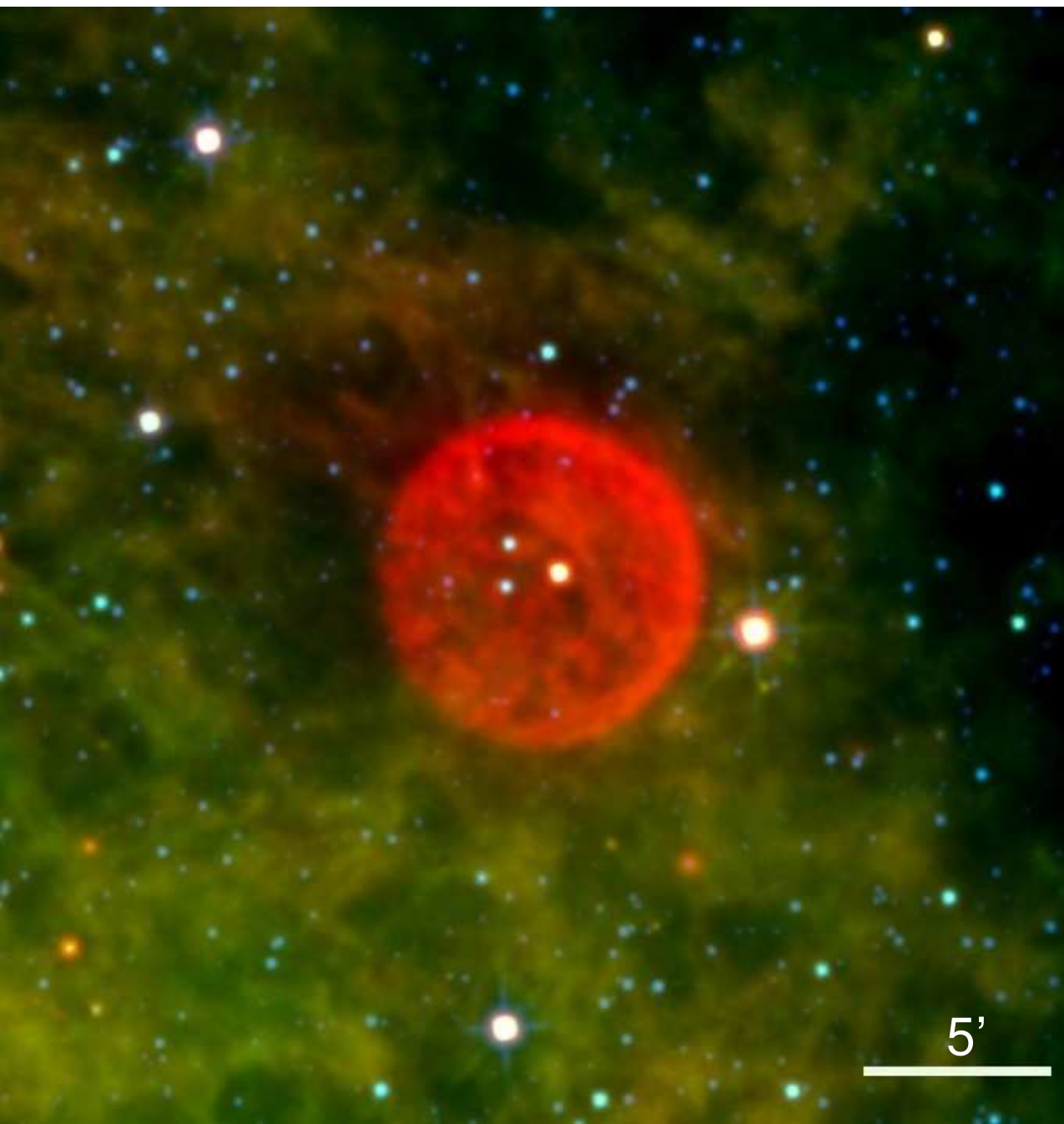}
\includegraphics[width=4.5cm]{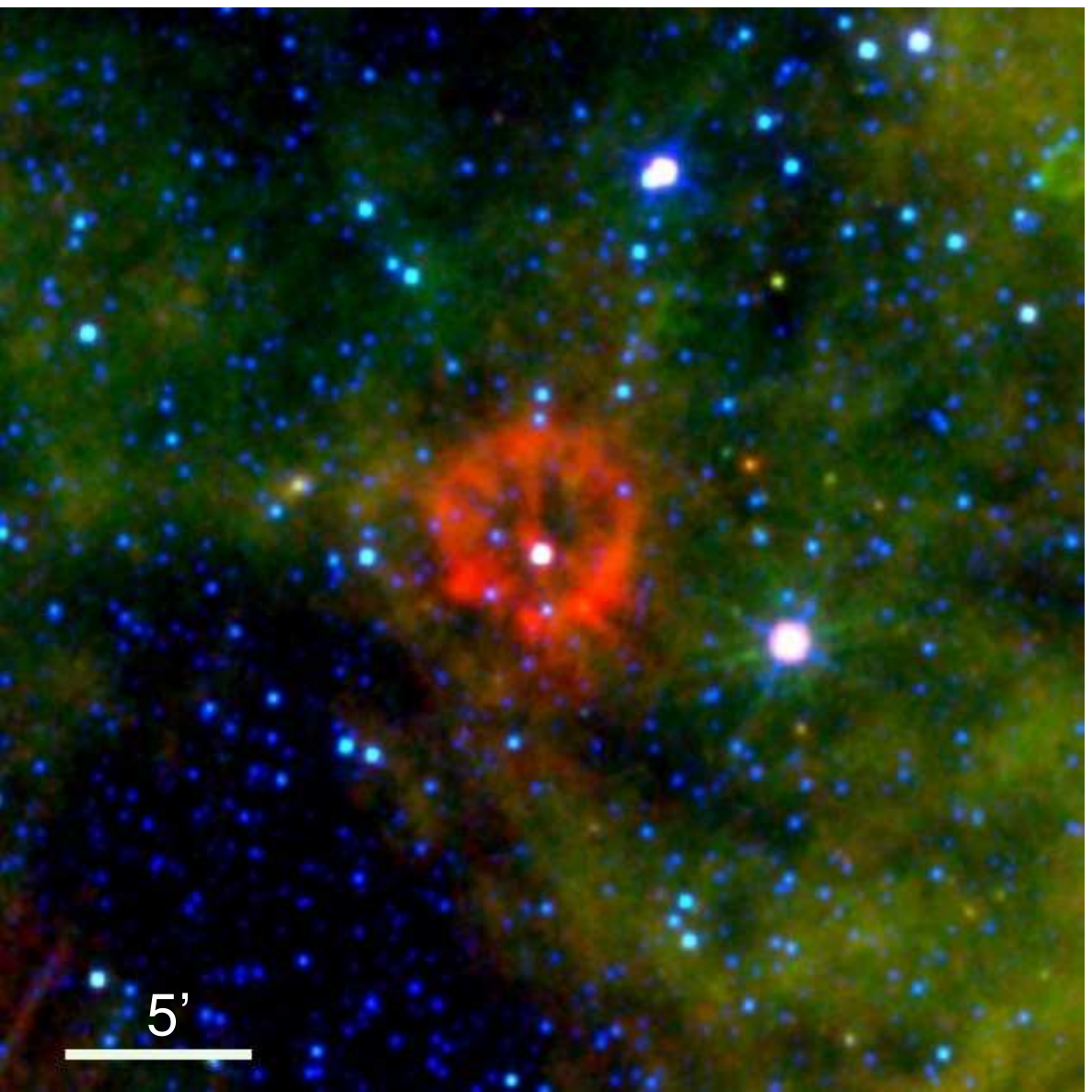}
\includegraphics[width=3.5cm]{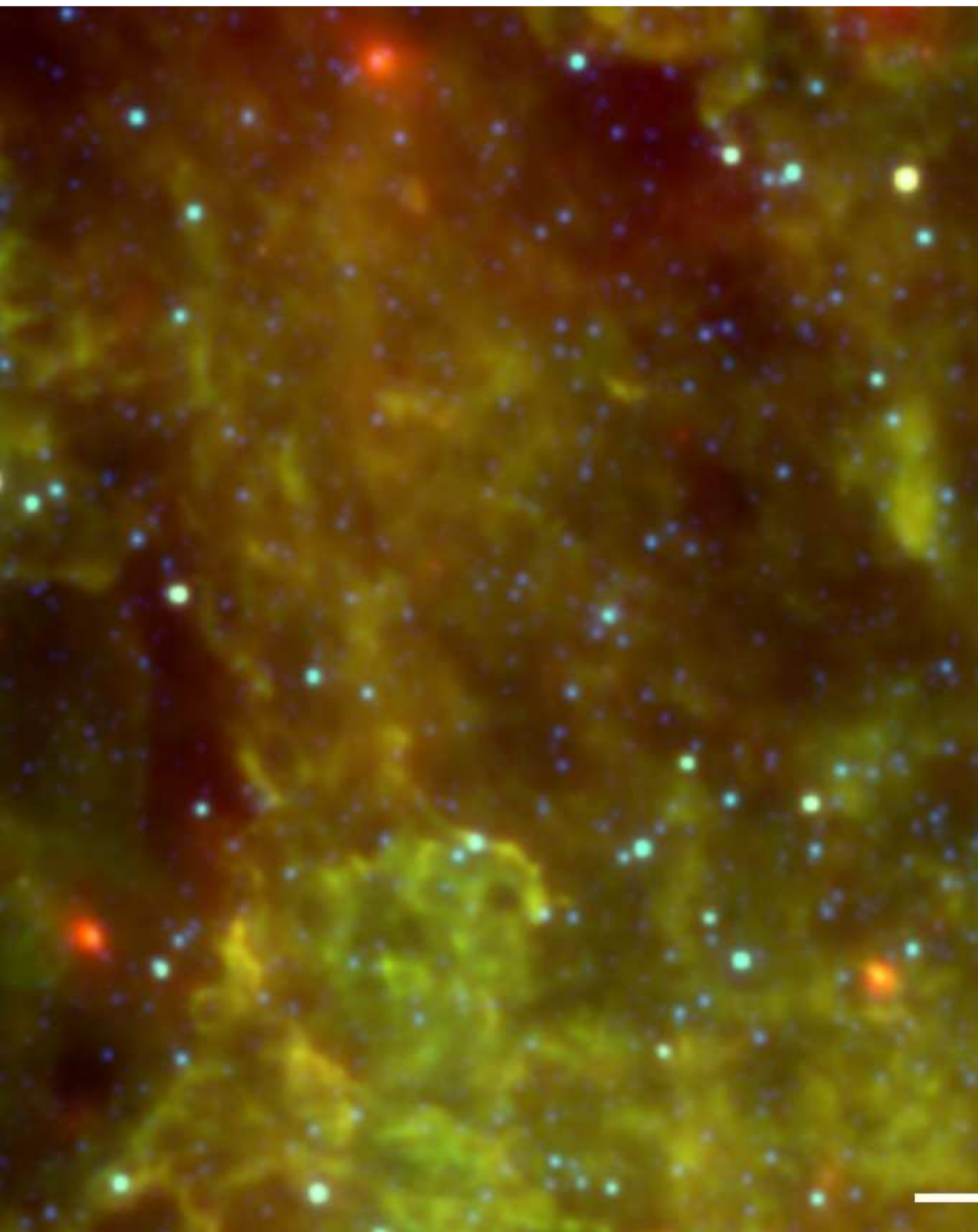}
\caption{Examples of the three stages in WR nebula morphologies: from left to right - bubble (WR16), clumpy phase (WR8), and mixed phase (WR35b). From Toal\'a et al. (in prep.).}
\label{Martin}
\end{center}
\end{figure}

IR is also useful in revealing details of particular objects. For example, a Herschel survey of LBVs undertaken at Li\`ege yielded as first result a characterization of the surroundings of WRAY 15-751 (Vamvatira-Nakou et al., submitted). IR photometry can only be explained if the star evolves at constant luminosity and dust grains are Fe-rich. Images also revealed the presence of a second shell, about 4 times larger than the previously known one, which most probably results from an older eruptive event. Considering both structures, there is about 0.075 M$_{\odot}$ of dust in the system. Ionized gas is responsible for several forbidden lines observed in the Herschel-PACS range (Fig. \ref{Chloi}), which allow a N/O abundance of about 7 times solar and a mass of ionized gas of 1--2 M$_{\odot}$ (20 times that of dust), to be derived.

Dust can be well studied in the IR, so this range may provide clues on where dust come from in galaxies. Two examples of such feedback were presented in the session: $\eta$\,Car and SN1987A. The latter was observed with Herschel at 100-500$\mu$m wavelengths, and 0.4-0.7 M$_{\odot}$ of dust was detected - mostly silicates and amorphous carbon (Matsuura et al. 2011). It is thought that this dust come from the explosion, but the role played by previous mass-loss episodes, in particular the LBV phase, is not yet clear. For example, about 0.12 M$_{\odot}$ of dust was detected, thanks to 30$\mu$m MiniTAO observations, in the famous LBV $\eta$\,Car. Up to 80\% of that dust belonged to the torus, hence may not be related to the big 1843 event.

\section{Wolf-Rayet stars}
Only a few percentage (4-6\%) of Wolf-Rayet stars displays surrounding nebulosities in the WISE survey, and most are found around WN stars (Wachter, in prep). 

The morphological classification scheme of WR nebulae proposed by Chu (1981) has been revised in this meeting by Guerrero et al. (Toal\'a et al, in prep.) using \emph{WISE} IR images and SDSS or Super Cosmos sky survey H$\alpha$ images for 35 nebulae associated with WRs. Two \emph{WISE} bands were particularly used: the one at 12$\mu$m, which encompasses PAH lines and lines of low excitation ions, and that at 22$\mu$m, to which thermal emission from dust and lines of He~{\sc i} as well as high excitation ions contribute. Three phases are defined. In the first one, WR nebulae appear as complete shells or bubbles. It corresponds to the star just entering the WR stage, when its powerful wind sweeps up the previous slow and dense winds (from e.g. LBV or RSG stages). The second phase is the clumpy phase. At that point, the nebulae display knots of gas and dust connected by partial shells and arcs. It corresponds to an age of a few 10$^4$\,yr, when instabilities break down the swept-up shell. The stellar motion through the ISM has an impact on the morphology, for example one-sided arc may be sometimes seen. Finally, the mixed nebular phase ends the cycle, with no definite morphology nor always a 1-to-1 correspondence between optical and IR images. It corresponds to the last stage, when the circumstellar nebula begins to dissolve into the ISM.  

\section{Studying the close environment of massive stars}
The close environment of massive stars is the ``missing link'' between the star itself and the large circumstellar features. It plays a key role in understanding the mass-loss, but it is also difficult to probe directly.

Emission lines arising in the wind and circumstellar material are a classical way to study this region, as well as near-IR excess linked to disk-like features. In this context, Be and B[e] stars are targets of choice, and surprises are frequent. For example, Graus et al. (2012) found three new sgB[e] in the SMC: they display typical spectra, with forbidden lines, but the line strengths as well as the IR excess appear reduced compared to usual objects of this class. It suggests that either the disks have less material or less dust than usual, or maybe that these stars are transitional objects. Another case intriguingly shows the opposite situation: CD$-$49$^{\circ}$3441 displays forbidden lines and appreciable IR excess, but is a main-sequence Be star away from any star-forming region (Lee and Chen 2009). A possibility may be that this star is in fact a weak B[e], rather than a classical Be. 

\begin{figure}
\centering
\includegraphics[width=12cm]{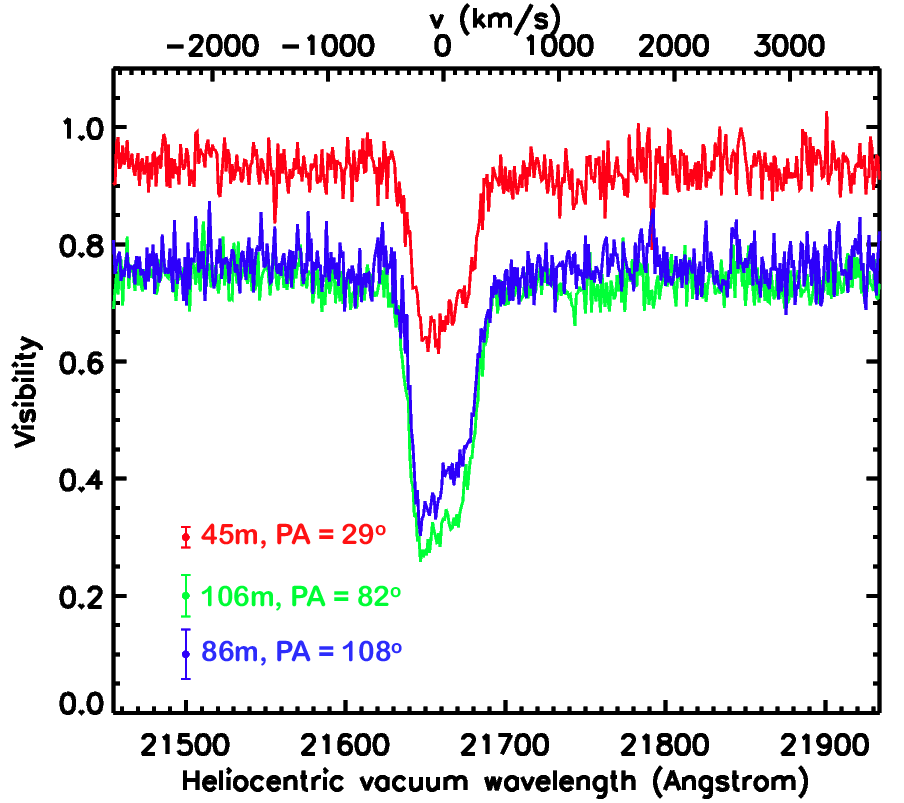}
\caption{
Visibility of HD 316285 as a function of wavelength for the three different telescope baselines measured with VLTI/AMBER. Notice the drop in visibility within the Bracket gamma line, indicating that the line is more extended than the neighboring K-band continuum. From Groh et al. (in prep.).}
\label{groh}
\end{figure}

The environment close to the star can also be studied, directly, by means of interferometry, which is usually performed at long wavelengths. Most optical/IR interferometric measurements rely on the measurements of ``visibilities'', which are directly linked to the size of the object. Recently, several massive stars, including nine LBVs, were observed with the VLTI (Groh et al., in prep.). Amongst these, HD316285 (Fig. \ref{groh}): the recorded visibilities implied a size of 0.002'' for the source of continuum radiation, and 0.004'' for the source of the Br$\gamma$ line. A CMFGEN fit to the spectrum yields a stellar model with which one can estimate the wind size in IR, and it agrees well with VLTI observations. The asymetric shape of the measured differential phases (red vs blue side) favors a prolate shape for the rotating star wind, but it could also be explained by clumps or binarity. Since the latter imply in time variability, a monitoring will be needed to ascertain the exact nature of the asymmetry.

\begin{figure}
\begin{center}
\includegraphics[width=10cm]{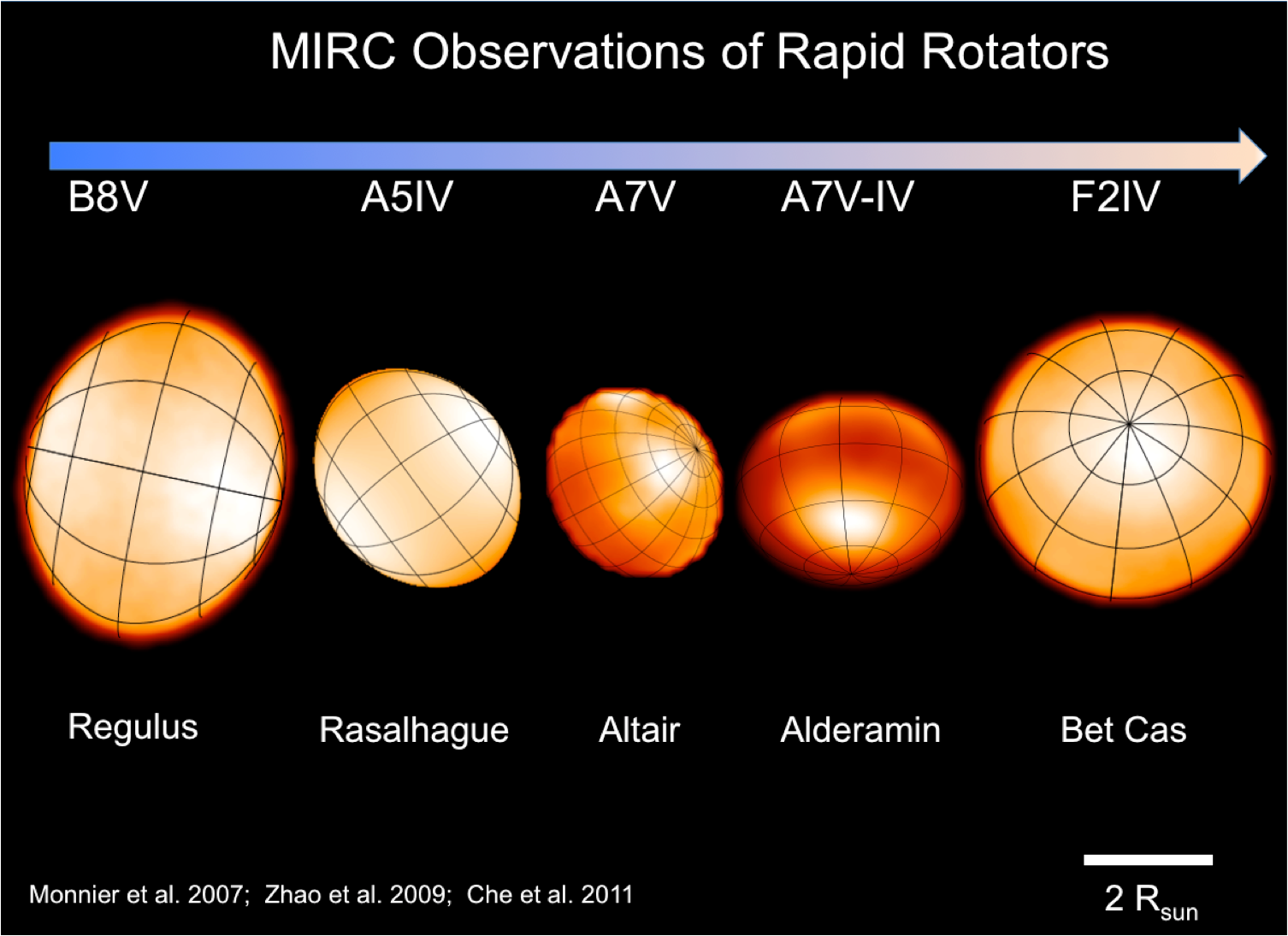}
\caption{Images of rapid rotators derived from MIRC interferometric measurements. Each star shows a bright pole and dark equator, which is caused by gravity darkening effect. Modeling of Regulus gives gravity darkening coefficient $\beta$ = 0.19, rather than 0.25 as von Zeipel predicted in 1924.}
\label{che}
\end{center}
\end{figure}

While visibilities provide valuable data, ``real'' images are always more impressive. Interferometric instruments such as Michigan Infrared Combiner (MIRC) and PIONIER are beginning to provide such data. MIRC was the first to image Altair (Monnier et al. 2007) and several other rapidly rotating stars as shown in Fig. \ref{che}. It has also imaged circumstellar disks and multi-object systems. For example, the disk contribution of $\delta$\,Sco was shown to remain stable during the periastron in 2011 (Che et al. 2012), and the mass-exchange in the $\beta$\,Lyr system can be clearly imaged (Zhao et al. 2008), as well as the 3 components of Algol (Baron et al. 2012) or the disk of the eclipsing companion of $\epsilon$\,Aur (Kloppenborg et al., 2010).  

\section{Conclusion}
This session has demonstrated the usefulness of studies in the IR in studying the environment of massive stars. Recent advances in this domain are notably provided by surveys, as they enable discovery of new objects to study, thereby improving the census of nebular features associated with hot stars. Furthermore, IR diagnostics unveil the properties of these neighbouring nebulosities: morphology, temperature, composition, density are the necessary keys paving the way of a better understanding of the mass-loss in massive stars.

\begin{acknowledgments}
YN acknowledges comments from Augusto Daminelli and support from FNRS and Prodex Herschel/XMM-Integral contracts. NLJC thanks FWO and Prodex-Herschel for financial support. JHG is supported by an Ambizione Fellowship of the Swiss National Science Foundation. CDL has been supported financially by grant NSC-101-2922-I-008-120 of the National Science Council of Taiwan. 

\end{acknowledgments}


\begin{thebibliography}{}


\bibitem[Baron et al. (2012)]{bar12}
     {Baron, F., et al.} 2012,
     \textit{ApJ} 752, 20

\bibitem[Blondin \& Koerwer (1998)]{Blondin98}
	{{Blondin}, J.~M., {Koerwer}, J.~F.} 1998,
	\textit{New Astronomy} 3, 571
	
\bibitem[Che et al. (2012)]{che12}
     {Che, X., et al.} 2012,
     \textit{ApJ} 757, 29

\bibitem[Chu (1981)]{chu81}
     {Chu, Y.-H.} 1981,
     \textit{ApJ}  249, 195

\bibitem[Clark et al. (2005)]{cl05}
     {Clark, J. S., Larionov, V. M., Arkharov, A.} 2005,
     \textit{A\&A} 435, 239

\bibitem[Comer{\'o}n \& Kaper (1998)]{Comeron98}
	{{Comeron}, F., {Kaper}, L.} 1998,
	\textit{A\&A} 338, 273

\bibitem[Cox et al. (2012)]{Cox12}
     {Cox, N., Kerschbaum, F., van Marle, A.-J., et al.} 2012,
     \textit{A\&A}  537, A35

\bibitem[Crowther et al. (2006)]{cro06}
     {Crowther, P. A., Lennon, D. J., Walborn, N. R.} 2006,
     \textit{A\&A} 446, 279

\bibitem[Dgani, Van Buren \& Noriega-Crespo (1996)]{Dgani96}
	{{Dgani}, R., {van Buren}, D., {Noriega-Crespo}, A.} 1996,
	\textit{ApJ} 461, 927

\bibitem[Graus et al. (2012)]{gra12}
     {Graus, A. S., Lamb, J. B., Oey, M. S.} 2012,
     \textit{ApJ} in press (arXiv:1208.5486) 

\bibitem[Groenewegen et al. (2011)]{gro11}
     {Groenewegen, M., et al.} 2011,
     \textit{A\&A} 526, A162

\bibitem[Gvaramadze et al. (2010)]{gva10}
     {Gvaramadze, V.V., Kniazev, A.Y., \& Fabrika, S.} 2010,
     \textit{MNRAS} 405, 1047

\bibitem[Gvaramadze et al. (2012)]{gva12}
     {Gvaramadze, V.V., et al.} 2012,
     \textit{MNRAS} 421, 3325

\bibitem[Haubois et al. (2009)]{hau09}
     {Haubois, X., et al.} 2009,
     \textit{A\&A} 508, 923

\bibitem[Huthoff \& Kaper (2002)]{Huthoff2002}
	{{Huthoff}, F., {Kaper}, L.} 2002,
	\textit{A\&A} 383, 999

\bibitem[Kervella et al. (2009)]{ker09}
     {Kervella, P., Verhoelst, T., Ridgway, S. T., Perrin, G., Lacour, S., Cami, J., Haubois, X.} 2009,
     \textit{A\&A} 504, 115

\bibitem[Kervella et al. (2011)]{ker11}
     {Kervella, P., Perrin, G., Chiavassa, A., Ridgway, S. T., Cami, J., Haubois, X., Verhoelst, T.} 2011,
     \textit{A\&A} 531, A117

\bibitem[Lee \& Chen (2009)]{lee09}
     {Lee, C.~D., \& Chen, W.~P.} 2009,
     \textit{ASPC} 404, 302

\bibitem[Kloppenborg et al. (2010)]{klo10}
     {Kloppenborg, B., et al.} 2010,
     \textit{Nature} 464, 870 

\bibitem[Matsuura et al. (2011)]{mat11}
     {Matsuura, M., et al.} 2011,
     \textit{Science} 333, 1258

\bibitem[Mizuno et al. (2010)]{miz10}
     {Mizuno, D.R., et al.} 2010,
     \textit{AJ} 139, 1542

\bibitem[Monnier et al. (2004)]{mon04}
     {Monnier, J.D. et al.} 2004,
     \textit{ApJ} 605, 436

\bibitem[Monnier et al. (2007)]{mon07}
     {Monnier, J. D.} 2007,
     \textit{Science} 317, 342

\bibitem[Noriega-Crespo, Van Buren \& Dgani (1997)]{NoriegaCrespo97}
	{{Noriega-Crespo}, A., {van Buren}, D., {Dgani}, R.} 1997,
	\textit{AJ} 113, 780

\bibitem[Peri et al. (2012)]{Peri12}
	{{Peri}, C.~S., {Benaglia}, P., {Brookes}, D.~P., {Stevens}, I.~R., 
	{Isequilla}, N.~L.} 2012,
	\textit{A\&A} 538, A108

\bibitem[Przybilla et al. (2010)]{prz10}
     {Przybilla, N., Firnstein, M., Nieva, M. F., Meynet, G., Maeder, A.} 2010,
     \textit{A\&A} 517, A38 

\bibitem[Smartt et al. (2002)]{sma02}
     {Smartt, S. J., Lennon, D. J., Kudritzki, R. P., Rosales, F., Ryans, R. S. I., Wright, N.} ,
     \textit{A\&A} 391, 979

\bibitem[Smith (2007)]{smi07}
     {Smith, N.} 2007,
     \textit{AJ} 133, 1034

\bibitem[Stencel et al. (1988)]{ste88}
     {Stencel, R. E., Pesce, J. E., Hagen Bauer, W.} 1988,
     \textit{AJ} 95, 141

\bibitem[Stencel et al. (1989)]{ste89}
     {Stencel, R. E., Pesce, J. E., Hagen Bauer, W.} 1989,
     \textit{AJ} 97, 1120

\bibitem[Stringfellow et al. (2012a)]{str12a}
     {Stringfellow, G.S., Gvaramadze, V.V., Beletsky, Y., \& Kniazev, A.Y.} 2012a,
     \textit{ASP Con Series} in press, arXiv:1112.2686

\bibitem[Stringfellow et al. (2012b)]{str12b}
     {Stringfellow, G.S., Gvaramadze, V.V., Beletsky, Y., \& Kniazev, A.Y.} 2012b,
     \textit{IAU Symp.} 282, 267

\bibitem[Van Buren \& McCray (1988)]{van88}
     {Van Buren, D., \& McCray, R.} 1988,
     \textit{ApJ} 329, L93

\bibitem[Wachter et al. (2010)]{wac10}
     {Wachter, S., Mauerhan, J.C., Van Dyk, S.D., Hoard, D.W., Kafka, S., \& Morris, P.W.} 2010,
     \textit{AJ} 139, 2330

\bibitem[Weaver (1977)]{Weaver77}
         {Weaver, R.P.} 1977,
	 \textit{Ph.D. Thesis Colorado University, Boulder} 

\bibitem[Wilkin (1996)]{Wilkin96}
	{{Wilkin}, F.~P.} 1996,
	\textit{ApJ Letters} 459, 31

\bibitem[Zhao et al. (2008)]{zha08}
     {Zhao, M., et al.} 2008,
     \textit{ApJ} 684, L95

\end{thebibliography}
\end{document}